\documentclass[aps,prl,showpacs,floats,twocolumn,floats,superscriptaddress,floatfix]{revtex4-1}
\usepackage{graphicx}
\usepackage{color}
\usepackage{amsmath}    % need for subequations
\usepackage{subcaption}
\usepackage{hyperref}   % use for hypertext links, including those to external documents and URLs
\hypersetup{
	colorlinks=true,
	allcolors=blue}
\usepackage{bm}
\usepackage{amssymb}
\usepackage{soul}

\newcommand{\jncaddress}{Jawaharlal Nehru Centre for Advanced Scientific Research, Bangalore 560064, India}

\begin{document}
\title{Inertial migration in pressure-driven channel flow: beyond the Segre-Silberberg pinch}
\author{Prateek Anand}\affiliation{\jncaddress}
\author{Ganesh Subramanian}\email{sganesh@jncasr.ac.in}\affiliation{\jncaddress}
%\vspace*{-10em}

\begin{abstract}
We examine theoretically the inertial migration of a neutrally buoyant rigid sphere in pressure-driven channel flow, accounting for its finite size relative to the channel width\,(the confinement ratio). For sufficiently large channel Reynolds numbers\,($Re_c$), a small but finite confinement ratio qualitatively alters the inertial lift velocity profiles obtained using a point-particle formulation. Finite size effects are shown to lead to new equilibria, in addition to the well known Segre-Silberberg pinch locations. Consequently, a sphere can migrate to either the near-wall Segre-Silberberg equilibria, or the new stable equilibria located closer to the channel centerline, depending on $Re_c$ and its initial position. Our findings are in accord with recent experiments and simulations, and have implications for passive sorting of particles based on size, shape and other physical characteristics, in microfluidic applications.
\end{abstract}
\maketitle 
Inertia-driven cross-stream migration of neutrally buoyant spheres in pipe flow, to an annular location between the centerline and walls, was first observed by Segre and Silberberg\cite{segre1961,segresilberberg1962a,segresilberberg1962b}, the location termed the Segre-Silberberg annulus\,(henceforth, SS-annulus or equilibria). Equilibria arising from inertial lift forces have since been exploited in a range of microfluidic applications\citep{dicarlo2007,dicarlo2009,dicarlo2014,dicarlo2017,dicarlo2020}. The first theoretical explanations of the phenomenon were for pressure-driven channel flow\,(the plane Poiseuille profile)\citep{holeal1974,vasseur1976}, and involved determining the inertial lift on a sphere for $Re_p, Re_c\ll1$, $Re_p$ and $Re_c$ being the particle and channel Reynolds numbers, respectively \cite{footnote1}. The pair of zero-crossings of the $O(Re_p)$ lift profile, symmetrically located about the centerline, corresponded to the SS-equilibria. The calculations were later extended to $Re_c \sim O(1)$ and larger\cite{hinch1989,asmolov1999}, with the SS equilibria starting at a location intermediate between the walls and centerline for $Re_c \ll 1$, and moving wallward with increasing $Re_c$. An analogous dependence on the Reynolds number was predicted for the SS-annulus in pipe flow\cite{matas2009}, pointing to the similar physics governing inertial migration in the two configurations. 

Later experiments\cite{matas2004}, while confirming the original observations\cite{segre1961,segresilberberg1962a,segresilberberg1962b}, revealed an additional inner annulus, this being the only equilibrium location beyond a certain $Re_c$\cite{footnote6}. The calculations above\citep{holeal1974,vasseur1976,hinch1989,asmolov1999} use a point-particle approximation, and predict only the pair of SS-equilibria in plane Poiseuille flow, and the SS-annulus alone in pipe flow\cite{matas2009}, regardless of $Re_c$. The inner annulus was therefore speculated to arise from finite-size effects\cite{matas2004}. Although initially regarded as a transient feature\cite{seki2017}, recent experiments\cite{nakayama2019} have confirmed the inner annulus to be a stable equilibrium, leading to the following migration scenario: all spherical particles focus onto the SS-annulus at low $Re_c$\,(Regime A); for $Re_c$ greater than a threshold, particles focus onto either the SS-annulus or the aforementioned inner annulus 
 depending on their radial distance from the centerline\,(Regime B); particles focus solely onto the inner annulus beyond a second threshold\,(Regime C). The threshold $Re_c$'s demarcating different regimes are observed to decrease with increasing $\lambda$, $\lambda$ being the confinement ratio defined as the ratio of the sphere radius $a$ to channel width $H$\,(or pipe radius). This scenario has been qualitatively confirmed by simulations\cite{shao2008,glowinski2021}, although the parameter ranges explored in the above studies are necessarily restricted.

In this letter, for the first time, we move beyond earlier point-particle formulations, and theoretically examine inertial migration in plane Poiseuille flow for small but finite $\lambda$, with $Re_c = V_\text{max}H/\nu$ being arbitrary; $V_\text{max}$ here is the centerline speed of plane Poiseuille flow, and $\nu$ the kinematic viscosity of the suspending fluid. $Re_p =\lambda^2 Re_c$ is assumed small, allowing analytical progress based on a leading order Stokesian approximation. The $O(\lambda Re_p)$ finite-size contribution is shown to qualitatively alter the inertial lift profiles obtained from a point-particle formulation\,($\lambda= 0$), for large $Re_c$, in a manner consistent with the recent studies above. Our calculations show that a new pair of stable equilibria, closer to the centerline, emerges beyond a threshold $Re_c$, even for $Re_p\ll1$. We provide a complete characterization of migration scenarios in the $\lambda\!-\!Re_c$ plane for plane Poiseuille flow.

\begin{figure}[!ht]
	\includegraphics[width=\columnwidth]{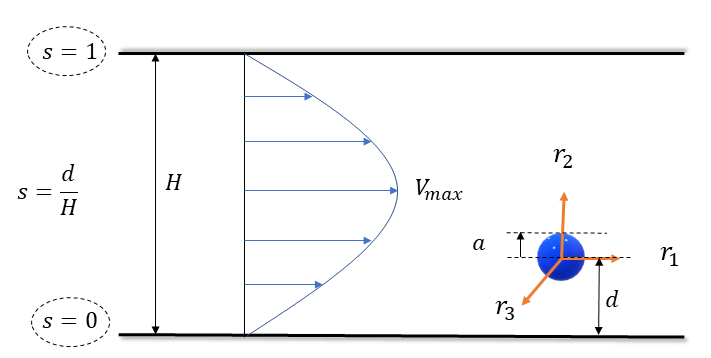}
	\caption{Neutrally buoyant sphere of radius $a$ in pressure-driven flow through a channel of width $H$.}
	\label{fig:schematic}
\end{figure}

For a neutrally buoyant rigid sphere in plane Poiseuille flow, at a non-dimensional distance of $\lambda^{-1}s$ from the lower wall\,(see Fig.\ref{fig:schematic}), use of the generalized reciprocal theorem leads to the following expression for the inertial lift velocity\cite{SM}:
\begin{widetext}
\begin{align}
	&V_p(s)=-Re_p\int_{V^F+V^P}\big[\bm{u}_\text{St}\cdot\big(\bm{u}_\text{str}\cdot\bm{\nabla U}^\infty+\bm{U}^\infty\cdot\bm{\nabla u}_\text{str}\big)dV+\lambda Re_p \Bigg[-\int_{V^\infty}\bm{u}_i\cdot\big(\bm{u}_{s,i}\cdot\bm{\nabla}\bm{u}_{s,i}\big)\,dV-\int_{V^\infty}\Big[\bm{u}_i\cdot\big(\bm{u}_{s,i}\cdot\nabla \bm{u}^\infty\nonumber\\
	&+\bm{u}^\infty\cdot\bm{\nabla u}_{s,i}\big)-\bm{u}_{\text{St},i}\cdot\big(\bm{u}_{\text{str},i}\cdot\bm{\nabla U}^\infty+\bm{U}^\infty\cdot\bm{\nabla u}_{\text{str},i}\big)\Big]\,dV+\int_{V^P}\bm{u}_{\text{St},i}\cdot\big(\bm{u}_{\text{str},i}\cdot\bm{\nabla U}^\infty+\bm{U}^\infty\cdot\bm{\nabla u}_{\text{str},i}\big)\,dV\Bigg],
	\label{eq:Ch4VpIntegralFinal}
\end{align}
\end{widetext}
for $Re_p$ small but finite, $V_p$ in being scaled with $V_\text{max}\lambda$.
 
The first integral in (\ref{eq:Ch4VpIntegralFinal}) is the point-particle contribution examined earlier\citep{holeal1974,vasseur1976,hinch1989,asmolov1999}, the domain of integration\,($V^F\!+\!V^P$) being the total volume contained between the channel walls. The dominant contribution to this integral comes from scales of $O(H)$, whence the finite sphere size may be neglected. Thus, $\bm{u}_\text{St}$ and $\bm{u}_\text{str}$ in the integrand are, respectively, the velocity fields due to a Stokeslet and a stresslet confined between the channel walls. The Stokeslet is oriented perpendicular to the walls, while the stresslet is proportional to the rate of strain tensor, associated with the plane Poiseuille flow, evaluated at the sphere location. This tensor is $\frac{1}{2}\beta(\bm{1}_1\bm{1}_2 + \bm{1}_2\bm{1}_1)$, with the Poiseuille flow given by $\bm{U}^\infty=(\beta r_2 + \gamma \lambda r_2^2)\bm{1}_1$ in a reference frame moving with the fluid velocity at the sphere center; here, $\bm{1}_1$ is a unit vector along the flow direction, $r_2$ the gradient coordinate relative to the sphere center, and $\beta=4(1-2s)$ and $\gamma=-4$ denote the shear rate and curvature of the plane Poiseuille profile. The dependence of the point-particle contribution on $H$ amounts to an $Re_c$-dependence in non-dimensional terms, so the first term in (\ref{eq:Ch4VpIntegralFinal}) is of the form $Re_p F_1(s,Re_c)$, with $F_1$ determined semi-analytically for $Re_c \ll 1$\citep{holeal1974,vasseur1976,anandJeffAvgd2022}, and numerically for $Re_c \gtrsim O(1)$\cite{hinch1989,asmolov1999,anandJeffAvgd2022}.

The integrals within square brackets, in (\ref{eq:Ch4VpIntegralFinal}), are the $O(\lambda Re_p)$ contributions. The dominant contributions to the first two integrals arise from scales of $O(a)$, so the channel walls may be neglected, the integration being over an unbounded fluid domain\,($V^\infty$) outside the sphere. Thus, $\bm{u}_i$ is the Stokesian velocity disturbance due to a sphere translating under a constant force normal to the walls, and $\bm{u}_{s,i}$ is the Stokesian disturbance due to a force-free torque-free sphere in an ambient plane Poiseuille flow, both in an unbounded fluid domain; $\bm{u}^\infty =(\beta r_2 + \lambda\gamma r_2^2 -\lambda\gamma/3)\bm{1}_1$ is the Poiseuille flow in a reference frame translating with the sphere, and differs from $\bm{U}^\infty$ above since the sphere translation speed includes a contribution\,($\lambda\gamma/3$) from the profile curvature\cite{kimkarrila,SM}. $\bm{u}_{\text{St},i}$ and  $\bm{u}_{\text{str},i}$ denote the unbounded-domain Stokeslet and the stresslet, respectively, the former given by the Oseen-Burgers tensor\cite{kimkarrila}; they differ from $\bm{u}_\text{St}$ and $\bm{u}_\text{str}$ above in not including the wall-image contributions \cite{lealbook,kimkarrila}. The third integral within brackets corrects for the inclusion of the sphere volume\,($V^P$) in the domain of integration of the point-particle integral.

The irrelevance of $H$ for the finite-size integrals implies that the expression within square brackets, in (\ref{eq:Ch4VpIntegralFinal}), is only a function of $s$. Further, the simple domain of integration\,($V^\infty$ or $V^P$) leads to this $s$-dependence being evaluable in closed form\cite{SM}, and (\ref{eq:Ch4VpIntegralFinal}) reduces to:
\begin{align}
	V_p(s)=Re_p\big[F_1(s,Re_c)+ \lambda \frac{1141(1-2s)}{216}\big],
	\label{eq:Ch4VpFinal}
\end{align}

\begin{figure*}
	\centering
	\begin{subfigure}[t]{\columnwidth}
	    \includegraphics[width=\columnwidth]{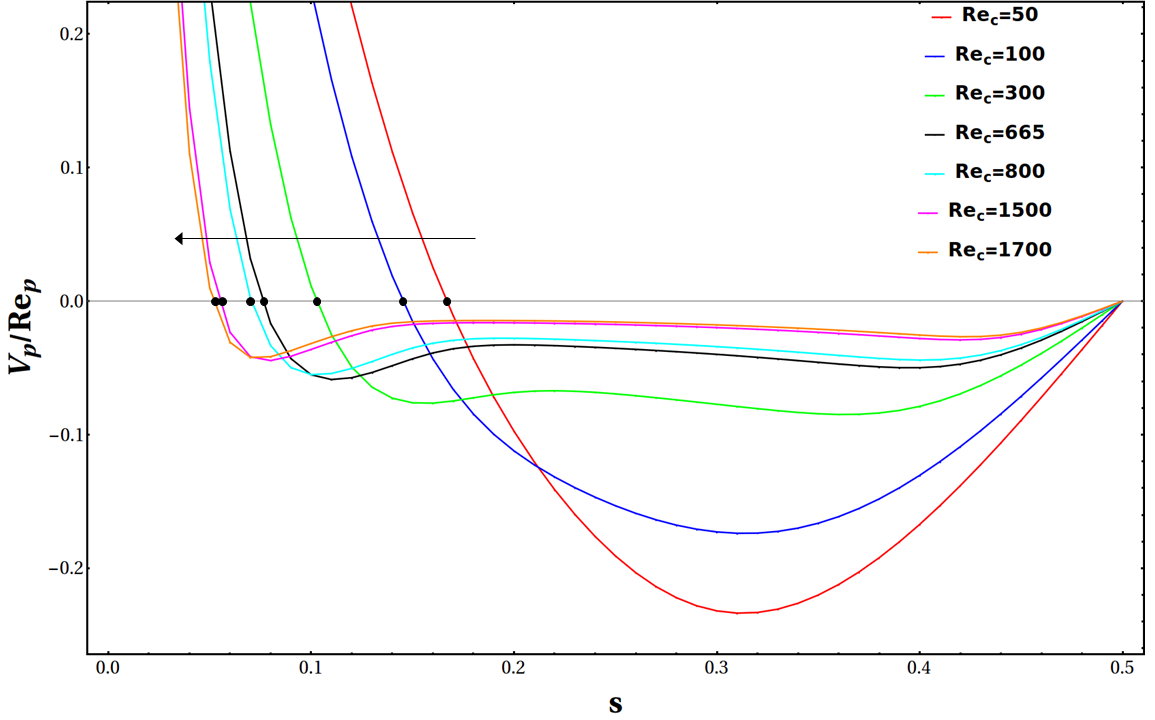}
	    \caption{$\lambda=0$ (point-particle)}
	\end{subfigure}
	\hfill
	\begin{subfigure}[t]{\columnwidth}
	    \includegraphics[width=\columnwidth]{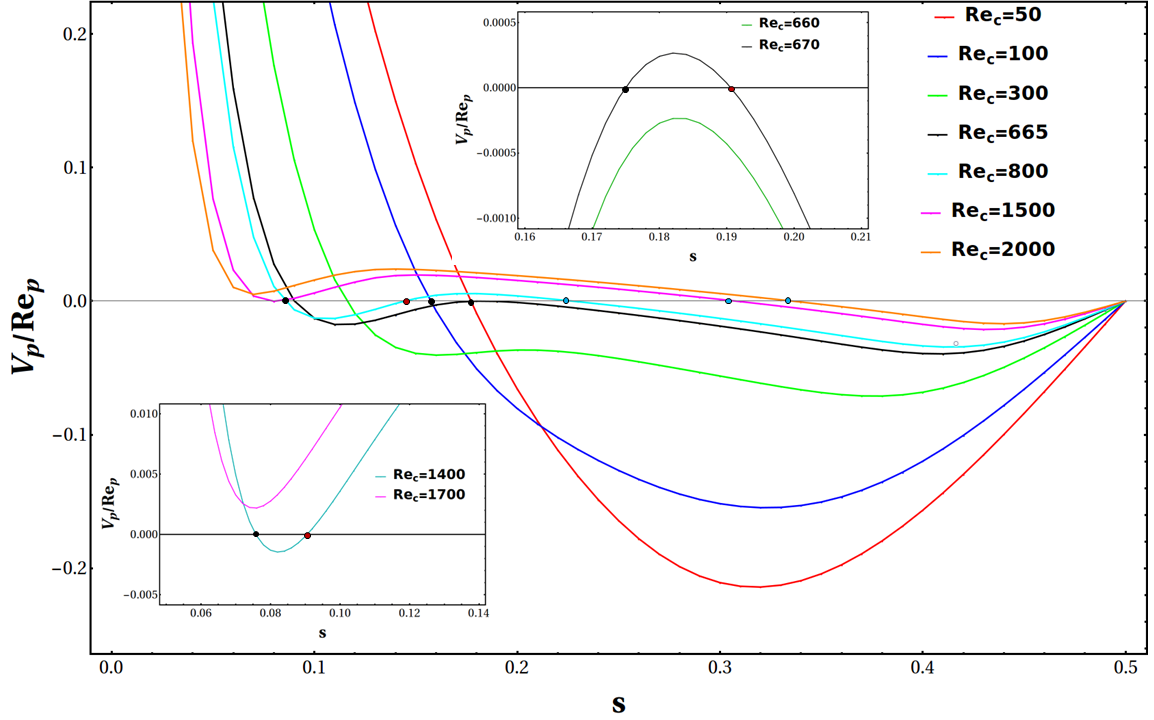}
	    \caption{$\lambda=0.01$}
	\end{subfigure}
	\caption{Inertial lift profiles in the lower half-channel for $Re_c\!\in\!(50,2000)$: (a) $\lambda=0$\,(point-particle); (b) $\lambda=0.01$; the two insets for $\lambda = 0.01$ provide a magnified view of the saddle-node bifurcations at $Re_{c_1}^\text{threshold} \approx 665$ and $Re_{c_2}^\text{threshold}\approx 1500$. The black, red and blue symbols denote the SS, unstable and stable equilibria, respectively. The arrow in (a) shows the movement of the SS equilibria with increasing $Re_c$.}
	\label{fig:Ch4LiftprofilesRecComparison}
	%\vspace*{3in}
\end{figure*}

$F_1(s,Re_c)$ in \eqref{eq:Ch4VpFinal} can be computed numerically for any $Re_c$ using a shooting method\cite{hinch1989,asmolov1999,anandJeffAvgd2022}. In Fig.\,\ref{fig:Ch4LiftprofilesRecComparison}a, the resulting (scaled)\,lift profiles are shown, for different $Re_c$'s, over the lower half-channel with $s\in[0,0.5]$\,(due to anti-symmetry about the centerline). In addition to the wallward movement of the lone zero-crossing\,(the SS-equilibrium), the magnitude of the lift at any fixed location, not in the neighbourhood of the wall\cite{footnote2}, decreases sharply with increasing $Re_c$\cite{asmolov1999}. This reflects the weakened particle-wall interaction when the walls recede beyond the inertial screening length of $O(HRe_c^{-\frac{1}{2}})$, owing to a more rapid decay of the disturbance velocity field at these distances. Apart from the overall decrease in magnitude, the shape of the profile also changes, with an intermediate concave-downward portion emerging for $Re_c \gtrsim 296$. An analogous scenario prevails for pipe flow, although the lift is smaller than that for channel flow at the same $Re_c$\cite{matas2009,nakayama2019}.

The changes in the point-particle contribution above imply that the finite-size term in \eqref{eq:Ch4VpFinal}, although $O(\lambda)$ smaller for $Re_c\ll1$, becomes comparable for sufficiently large $Re_c$. This is seen in Fig.\, \ref{fig:Ch4LiftprofilesRecComparison}b which shows the lift  profiles, for $\lambda = 0.01$, for the same $Re_c$'s as in Fig.\,\ref{fig:Ch4LiftprofilesRecComparison}a. For $Re_c = 50$, the lift profile and the SS-equilibrium are only marginally affected. In contrast, for $Re_c=Re_{c_1}^\text{threshold}$ ($\approx 665$ for $\lambda=0.01$), while the SS-equilibrium (expectedly)\,has moved closer to the walls, a pair of stable and unstable equilibria appear between it and the centerline via a saddle-node bifurcation; the unstable equilibrium demarcating the basins of attraction of the SS equilibrium and the inner stable equilibrium. The bifurcation arises due to finite-size effects causing the region of negative curvature, in the point-particle profile, to cross the zero-lift line\,(upper inset in Fig.\,\ref{fig:Ch4LiftprofilesRecComparison}b).  As $Re_c$ increases to $800$, the unstable equilibrium moves towards the SS equilibrium even as both move wallward, while the inner stable equilibrium moves towards the centerline. A second saddle-node bifurcation at $Re_c=Re_{c_2}^\text{threshold}$ ($\approx1500$ for $\lambda=0.01$) leads to the near-center stable equilibrium being the only one in the half-channel for larger $Re_c$\,(lower inset in Fig.\,\ref{fig:Ch4LiftprofilesRecComparison}b). Note that for $Re_c\in(50,2000)$ as in Fig.\,\ref{fig:Ch4LiftprofilesRecComparison}b, and $\lambda=0.01$, $Re_p\in($0.005,0.2$)$, consistent with the theoretical assumption of weak fluid inertial effects on scales of $O(a)$.

\begin{figure*}
	\centering
	\begin{subfigure}[b]{\columnwidth}
	    \includegraphics[width=\columnwidth]{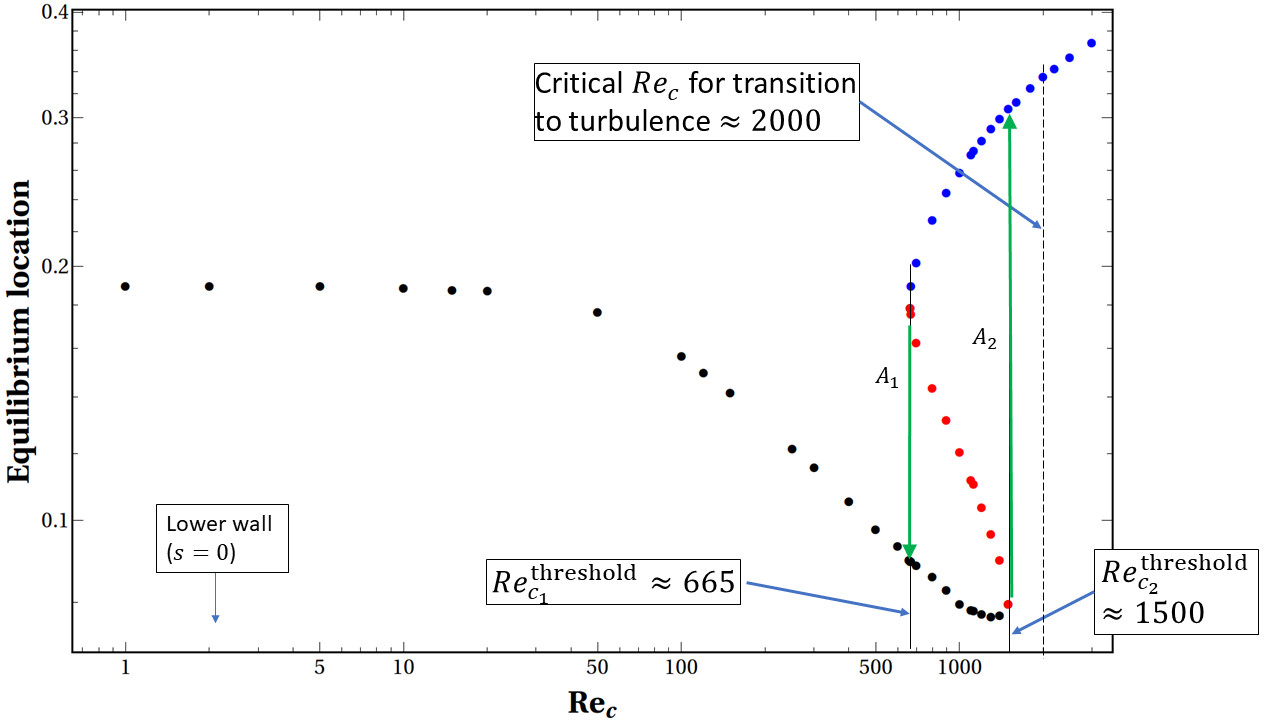}
	    \caption{$\lambda=0.01$}
	\end{subfigure}
	\hfill
	\begin{subfigure}[b]{\columnwidth}
	    \includegraphics[width=\columnwidth]{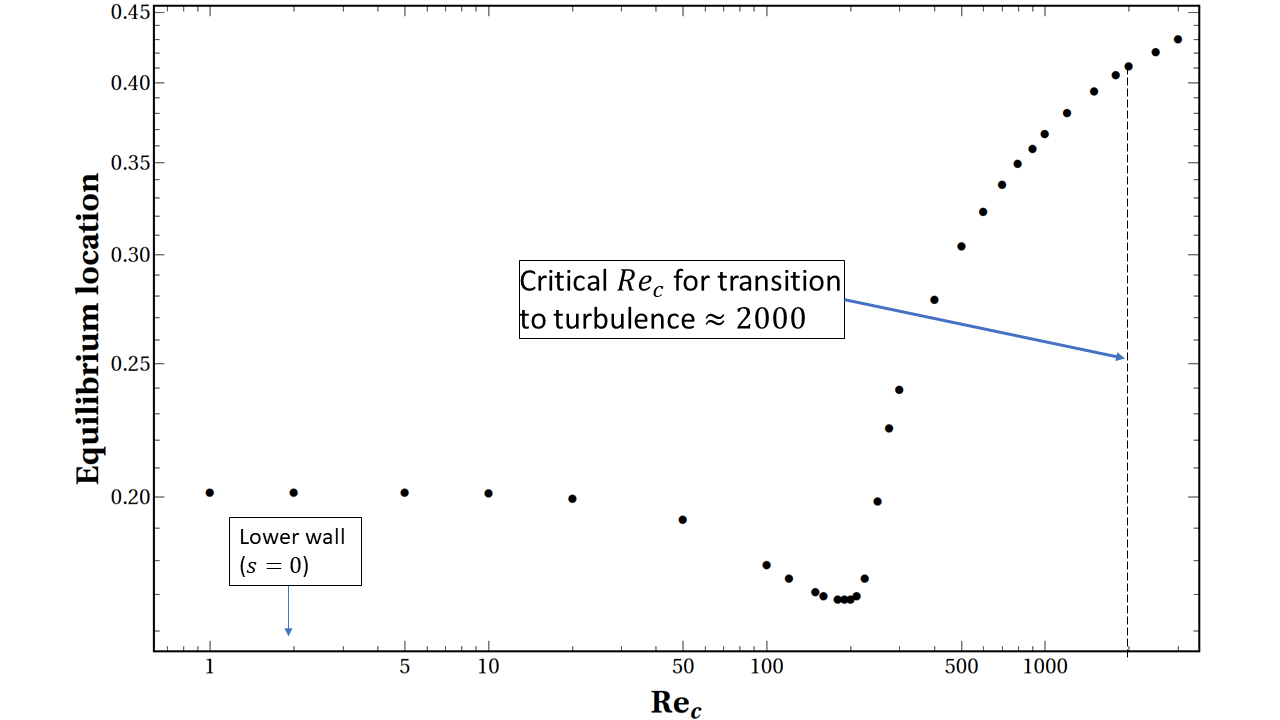}
	    \caption{$\lambda=0.025$}
	\end{subfigure}
	\caption{Inertial equilibrium loci for (a)$\lambda = 0.01$ and (b)$\lambda = 0.025$; black, blue and red dots denote the SS, and the inner stable and unstable equilibria, respectively. The region of multiple equilibria in (a), for $Re_c\in(665,1500)$, leads to hysteretic jumps in the equilibrium location marked by the vertical arrows $A_1$\,($s=0.179\rightarrow0.09$) and $A_2$($s=0.08\rightarrow0.31$). Vertical dashed lines in (a) and (b) denote the laminar-turbulent transition.}
	\label{fig:Ch4eqmLoci}
\end{figure*}

%Thus, for the chosen $\lambda$, one obtains only the SS equilibrium for $Re_c \lesssim 665$\,(starting at $s\approx0.182$ for $Re_c \ll 1$, and moving to smaller $s$), only the near-center equilibrium induced by finite-size effects for $Re_c \gtrsim 1500$\,(at $s\approx 0.18$ for $Re_c \approx 1500$, and moving to larger $s$), and both these stable equilibria\,(separated by an intervening unstable one) for $Re_c\in(665,1500)$. 

Fig.\,\ref{fig:Ch4eqmLoci}a plots the equilibrium loci identified above, for $\lambda = 0.01$, as a function of $Re_c$. The SS-branch is seen to start at $s\approx0.182$ for $Re_c \ll 1$, moving to smaller $s$ with increasing $Re_c$. The inner stable equilibrium emerges discontinuously at $s\approx 0.18$ for $Re_c \approx 665$, moving to larger $s$ thereafter\,(the SS-equilibrium is at $s \approx 0.09$ for this $Re_c$). The loci of the SS and the inner equilibria are shown as sequences of black and blue dots, respectively, with the unstable equilibrium locus connecting the two shown as a sequence of red dots. The fold that develops in the interval $Re_c\in(665,1500)$, bracketed by the two saddle-node bifurcations, implies a hysteretic behavior in an experiment \cite{footnote3}. A quasi-static protocol of increasing flow rate will lead to spheres remaining at the SS-equilibrium until $Re_c \approx 1500$, at which point they will jump onto the new stable branch closer to the centerline. In contrast, along a path of decreasing flow rate, spheres will remain at the inner stable equilibrium down to $Re_c \approx 665$, before jumping back to the SS-branch. 

A behavior analogous to that in Fig.\,\ref{fig:Ch4eqmLoci}a occurs for $\lambda$ less than $0.01$, with the pair of $Re_c$-thresholds increasing with decreasing $\lambda$. However, the equilibrium loci undergo a qualitative change as $\lambda$ increases. To see this, note that the SS-equilibrium, in the point-particle framework, emerges from a balance between an $O(\beta\gamma)$ curvature-induced contribution driving migration towards higher shear rates\,(that is, away from the centerline), and an $O(\beta^2)$ wall-induced repulsion. Both contributions arise due to inertial forces acting on scales of $O(H)$ for $Re_c \ll 1$\cite{anandJeffAvgd2022}, and decrease with increasing $Re_c$. The $O(\beta^2)$ contribution decreases faster, leading to the wallward movement of the SS-equilibrium. At $O(\lambda)$, there arises an additional curvature-induced contribution on scales of $O(a)$, and that drives migration towards the centerline \cite{footnote4}. The opposing signs of the point-particle and finite-size curvature-induced contributions weakens the wallward movement\,(with increasing $Re_c$) of the SS-equilibrium for larger $\lambda$. The profound effect of this weakening may be seen from Fig.\,\ref{fig:Ch4eqmLoci}b which shows the equilibrium locus for $\lambda = 0.025$. The region of multiple equilibria is now absent, with the SS-equilibrium smoothly transitioning from an initial wallward movement, to a movement towards the centerline, across $Re_c\approx 200$.

\begin{figure}[!ht]
	\includegraphics[width=\columnwidth]{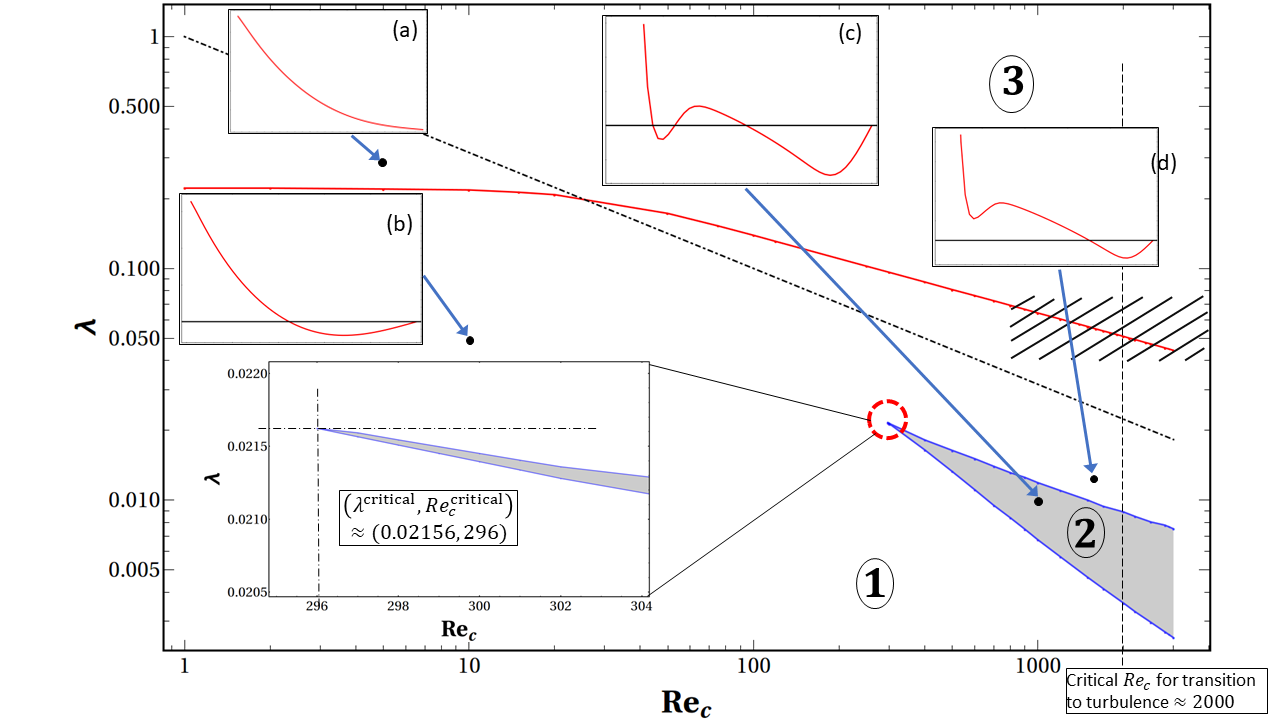}
	\caption{Migration scenarios in the $\lambda-Re_c$ plane. Inertial lift profiles for canonical $(\lambda,Re_c)$ pairs (highlighted by black dots) are shown\,(see text for details). The hatched region shows parameter ranges covered in earlier studies \cite{matas2004,seki2017,nakayama2019,shao2008,glowinski2021}.}
	\label{fig:Ch4PhaseDiag}
\end{figure}

By identifying the equilibrium loci as a function of $Re_c$, for different $\lambda$, a `phase diagram' of migration scenarios in the $\lambda-Re_c$ plane may be constructed as in Fig.\,\ref{fig:Ch4PhaseDiag}. The figure highlights the existence of three distinct regions. Region \textcircled{1}, corresponding to the area below the red curve and outside the (gray)\,shaded region, contains lift profiles with a single stable off-center equilibrium in the half-channel. Region \textcircled{2}, corresponding to the shaded region, contains lift profiles with two stable equilibria, separated by an intervening unstable one, in the half-channel. The upper and lower boundaries of this region are determined by the pair of turning points on the equilibrium locus, corresponding to saddle-node bifurcations - these were identified in Fig \ref{fig:Ch4eqmLoci}a for $\lambda = 0.01$. The two boundaries end in a cusp for the fold bifurcation under consideration \cite{zeeman1976,pulkit2021}, corresponding to $(\lambda^\text{critical}, Re_c^\text{critical})\equiv(0.0216,296)$ in Fig.\,\ref{fig:Ch4PhaseDiag}; see top right inset. Along either a vertical or a horizontal line, the latter corresponding to an experimental path of changing flow rate, Region \textcircled{2} mediates a discontinuous transition from the SS-equilibrium to the inner stable equilibrium closer to the centerline. Region \textcircled{3}, above the red curve, includes lift profiles with the centerline as the only stable equilibrium. Note that the centerline is always an equilibrium by symmetry, albeit an unstable one in Regions \textcircled{1} and \textcircled{2}. Insets in Fig.\,\ref{fig:Ch4PhaseDiag} show lift profiles for the following $(\lambda,Re_c)$ pairs: $(0.3,5)$; (b) $(0.05,10)$; (c) $(0.01,1000)$; (d) $(0.015,1500)$, which are consistent with the aforementioned features of Regions \textcircled{1}-\textcircled{3}. 

The black dot-dashed line in Fig.\,\ref{fig:Ch4PhaseDiag} corresponds to $Re_p=1$, and may be regarded as a rough threshold above which the present theoretical results may no longer be valid. This implies, for instance, that our results may not be quantitatively accurate beyond $Re_c=100$ at $\lambda=0.1$. Importantly, the region of multiple equilibria and the associated hysteretic transitions, predicted here for the first time, lie well within this threshold. A second factor that limits the observability of Region \textcircled{2} is the laminar-turbulence transition. Although plane Poiseuille flow is predicted to become linearly unstable at $Re_c=11544$ \cite{orszag1971}, experiments show a nonlinear subcritical transition to turbulence at a much lower $Re_c \sim O(2000)$\citep{carlson1982,nishioka1985}. This subcritical transition is shown as vertical dashed lines in both Figs.\,\ref{fig:Ch4eqmLoci} and \ref{fig:Ch4PhaseDiag}. The emergence of the region of multiple equilibria in the latter figure clearly occurs well before the transition threshold. 

While the migration pattern for a fixed $\lambda$, implied by Fig.\,\ref{fig:Ch4PhaseDiag}, is in qualitative agreement with studies quoted at the beginning\cite{matas2004,seki2017,nakayama2019,shao2008,glowinski2021}, the inner annulus in these studies is observed for higher $\lambda\,(\gtrsim 0.05)$ and for $Re_p \gtrsim O(1)$ - see hatched region in Fig.\,\ref{fig:Ch4PhaseDiag}. The absence of multiple equilibria for smaller $\lambda$ is very likely due to the development length, needed for a steady particle distribution, being larger than the pipe length used in the experiments. For $Re_c$ fixed, this length scales as $O(\lambda^{-3})$\citep{matas2004}, increasing rapidly with decreasing particle size. Notwithstanding differences between the pipe and channel geometries, experiments with longer pipes should lead to the hatched region in Fig.\,\ref{fig:Ch4PhaseDiag} extending down to smaller $\lambda$. There remains the provocative question of how the secondary finite-$Re_p$ region of multiple equilibria, identified in the said studies, connects to the theoretically identified small-$Re_p$ region\,(Region \textcircled{2} in Fig.\,\ref{fig:Ch4PhaseDiag}).

%\textcolor{red} {It is worth noting that the experiments of , and the numerical simulations of \cite{shao2008,glowinski2021}, that predict the inner annulus, were performed for $Re_p\gtrsim O(1)$. While quantitative agreement would require use of smaller particles in the experiments, 

%In this letter, we have characterized the effects of finite particle size on the lift velocity of a neutrally buoyant sphere freely suspended in plane Poiseuille flow. We have shown, for the first time, that the inclusion of finite-size contribution allows one to go beyond the original Segre-Silberberg picture, and obtain a pair of new stable equilibria, closer to the centerline. For $\lambda$'s below a certain threshold ($\lambda^\text{critical}\approx0.0215$), these new equilibria arise via a saddle-node bifurcation, in turn lending a hysteresis character to the evolution of the inertial equilibrium with changing $Re_c$. The loci of the equilibria obtained, with changing $Re_c$, qualitatively explain the experimental observations of \cite{matas2004,nakayama2019}. A phase diagram summarizing the loci is plotted in the $\lambda-Re_c$ plane. An obvious  advantage of such a phase diagram is that it predicts the effects of the finite particle size for a wide range of $\lambda$'s and $Re_c$'s which can be helpful in the rational design of future experiments; this is unlike the earlier numerical simulations which were restricted to certain selected $\lambda$'s and $Re_c$'s. 

Apart from the fundamental significance of our findings, in terms of enriching the inertial migration landscape, and providing an explanation for recent experiments and computations, Fig.\,\ref{fig:Ch4PhaseDiag} may be leveraged towards passive sorting in microfluidic applications. The simplest scenario pertains to separating spheres of two different sizes, corresponding to confinement ratios $\lambda_1$ and $\lambda_2$\,($\lambda_2 > \lambda_1$). An experimental protocol of changing flow rate\,($Re_c$) for a bi-disperse suspension, with particles of the aforementioned sizes, would appear as a pair of horizontal lines in Fig.\,\ref{fig:Ch4PhaseDiag}, the upper one corresponding to $\lambda_2$. With increasing flow rate, separation would be achieved at an $Re_c$ when the point on the $\lambda_2$-line is above Region \textcircled{2}\,(after crossing it to the right), with the one on the $\lambda_1$-line directly below. At this $Re_c$, larger spheres would focus onto the pair of near-centerline stable equilibria, with the smaller ones focusing onto the near-wall SS-equilibria. If the point on the $\lambda_1$-line lies within Region \textcircled{2}, rather than below it, as would be the case when $\lambda_2/\lambda_1$ is not far from unity, partial separation will be achieved owing to smaller spheres focusing onto both the SS and inner equilibria\,(the relative fractions determined by the pair of unstable equilibria).

The implications of the near-centerline stable equilibria found here go well beyond the size-sorting scenario above. The dependence of the interval of existence of these equilibria, on inertial forces in a region of order the sphere size, implies a generic sensitivity of the finite-size contributions to the detailed characteristics of the suspended microstructure. Thus, for anisotropic particles such as spheroids or ellipsoids, the threshold Reynolds numbers\,($Re_{c_1}^\text{threshold}$, $Re_{c_2}^\text{threshold}$) that characterize the region of multiple equilibria are expected to be functions of the particle aspect ratio(s). In contrast, the time-averaged inertial lift for spheroids with order unity aspect ratios, calculated within a point-particle framework for plane Poiseuille flow, may be shown to yield SS-equilibria identical to those for spheres\cite{anandJeffAvgd2022}; the spheroid aspect ratio only affecting the magnitude of the point-particle lift profile, not the equilibrium locations. The aspect-ratio-dependence expected for the near-centerline equilibria will therefore be crucial for shape-sorting protocols in microfluidic applications\cite{masaeli2012}. Along similar lines, there will be a dependence on the viscosity ratio for drops, allowing, in principle, for separation of weakly deformed drops based on both size and viscosity ratio differences\cite{footnote5}. Analogous remarks apply to other elastic microstructures such as vesicles, capsules or red blood cells. Migration phase diagrams for these cases will have at least one additional axis - this axis could be the appropriate shape parameter for anisotropic particles\,(aspect ratio for spheroids) or the viscosity ratio for drops. In the latter case, the degree of deformability, as characterized by the Capillary number, offers an additional degree of freedom. It would be of interest, in future, to quantitatively determine these phase diagrams, allowing for rational design of passive sorting protocols. 
 
\begin{acknowledgments}
\end{acknowledgments}

\clearpage
\onecolumngrid

\begin{center}
  \textbf{\large Supplemental material}\\[.2cm]
%   (Dated: \today)\\[1cm]
\end{center}

\setcounter{equation}{0}
\setcounter{figure}{0}
\setcounter{table}{0}
\setcounter{page}{1}
\renewcommand{\theequation}{S\arabic{equation}}
\renewcommand{\thefigure}{S\arabic{figure}}
\renewcommand{\bibnumfmt}[1]{[S#1]}
\renewcommand{\citenumfont}[1]{S#1}

Following \cite{holeal1974S} and \cite{anandJeffAvgd2022S}, one may use a generalized reciprocal theorem formulation to derive a formal expression for the inertial lift velocity of a neutrally buoyant sphere, in an ambient plane Poiseuille flow, for arbitrary $Re_p$. In the limit $Re_p \ll 1$, the non-dimensional inertial lift is $O(Re_p)$, being given by:
\begin{align}
	V_p=-Re_p\int_{V^F}\bm{u}\cdot\left(\bm{u}_s\cdot\bm{\nabla u}_s+\bm{u}_s\cdot\bm{\nabla u}^\infty+\bm{u}^\infty\cdot\bm{\nabla u}_s\right) dV. \label{eq:Ch4RTliftvelS}
\end{align}
where $V_p$ is scaled by $V_\text{max}a/H$ or $V_\text{max}\lambda$. 

The actual problem in the reciprocal theorem framework corresponds to the one of interest, that is, a neutrally buoyant sphere freely moving in wall-bounded plane Poiseuille flow for $Re_p$ and $\lambda$ small but finite, with $Re_c = \lambda^{-2}Re_p$ being arbitrary. For purposes of calculating the inertial lift to $O(Re_p)$, the disturbance velocity field in the actual problem may be replaced by its Stokesian approximation.  Thus, $\bm{u}_s$, in the inertial acceleration terms in (\ref{eq:Ch4RTliftvelS}), is the Stokesian disturbance field due to a force-free and torque-free sphere translating with $\bm{U}_p$ in an ambient plane Poiseuille flow. In a reference frame moving with the sphere center, the latter flow is given by $\bm{u}^\infty=(\alpha+\beta r_2+\lambda\gamma r_2^2)\bm{1}_1-\bm{U}_p$ where, with the sphere at a distance $\lambda^{-1}s$\,(in units of $a$) from the lower wall, one has $\alpha=4 \lambda^{-1} s(1-s)$, $\beta=4(1-2s)$ and $\gamma=-4$. Here, $\alpha\bm{1}_1-\bm{U}_p$ and $\beta$ are the ambient slip and shear rate at the sphere center, the latter varying linearly across the channel and equalling zero at the centerline\,($s=0.5$); $\gamma$ denotes the constant curvature of the Poiseuille profile. $\bm{U}_p$ is determined using the force-free constraint in Faxen's law for translation. This leads to $\bm{U}_p=(\alpha+\lambda\gamma/3)\bm{1}_1$ and, as a result,  the ambient flow in the said reference frame takes the form $\bm{u}^\infty=(\beta r_2+\lambda\gamma r_2^2-\lambda\gamma/3)\bm{1}_1$. The test problem in the reciprocal theorem framework corresponds to the Stokesian translation of a sphere in an otherwise quiescent fluid confined between parallel walls\,(ones that bound the Poiseuille flow in the actual problem), under a constant force acting along the gradient direction. The test disturbance field $\bm{u}$ multiplies the inertial acceleration terms involving $\bm{u}_s$ in (\ref{eq:Ch4RTliftvelS}).

We now examine the length scales that contribute dominantly to the integral in \eqref{eq:Ch4RTliftvelS}, beginning with the limit $Re_c\ll1$, when the inertial screening length ($HRe_c^{-1/2}$) is much larger than the channel width, and therefore, irrelevant. The dominant contributions to the volume integral in this limit may arise from either scales of $O(a)$\,(the inner region) or those of $O(H)$\,(the outer region). In order to assess their relative magnitudes, we consider the intermediate asymptotic interval $1\ll r\ll\lambda^{-1}$\,($r$ is measured in units of $a$) where both the finite size of the sphere and wall-induced image contributions may be neglected at leading order. For $r$ in this interval, $\bm{u}\sim 1/r$ corresponding to the farfield Stokeslet, and $\bm{u}_s\sim \beta/r^2+\gamma\lambda/r^3$ corresponding to the farfield stresslet and force-quadrupole contributions associated with the linear and quadratic ambient flow components, respectively. Using these forms along with $\bm{u}^\infty\sim \beta r + \gamma\lambda r^2$, one obtains the following estimates for parts of the integrand involving the linear and nonlinear components of the inertial acceleration:
\begin{itemize}
	\item $\bm{u}\cdot\left(\bm{u}_s\cdot\bm{\nabla u}^\infty+\bm{u}^\infty\cdot\bm{\nabla u}_s\right)\sim \dfrac{\beta^2}{r^3}+\dfrac{\lambda\beta\gamma}{r^2}+\dfrac{\lambda\gamma\beta}{r^4}+\dfrac{\gamma^2}{r^3}\,\,\,\,$ (linear),
	\item $\bm{u}\cdot(\bm{u}_s\cdot\bm{\nabla} \bm{u}_s)\sim \dfrac{\beta^2}{r^6}+\dfrac{\lambda\beta\gamma}{r^7}+\dfrac{\lambda^2\gamma^2}{r^8}\,\,\,\,\,\,\,$ (nonlinear).
\end{itemize}
Using $dV\sim O(r^2 dr)$, one obtains the following estimates for contributions to the lift velocity integral:
\begin{subequations}
\begin{align}
V_p^\text{linear}&\sim Re_p\int^r dr' \left( \,\frac{\beta^2}{r'}+\lambda\beta\gamma\,+\,\frac{\lambda\gamma\beta}{r'^2}+\,\frac{\lambda^2\gamma^2}{r'} \right), \nonumber\\
&\sim Re_p\left(\beta^2\ln r+\lambda\beta\gamma r\,+\,\frac{\lambda\gamma\beta}{r}+\lambda^2\gamma^2\ln r \right), \\
V_p^\text{non-linear}&\sim Re_p\int^r dr' \left(\,\frac{\beta^2}{r'^4}+\,\frac{\lambda\beta\gamma}{r'^5}+\,\frac{\lambda^2\gamma^2}{r'^6}\right), \nonumber\\
&\sim Re_p \left(\,\frac{\beta^2}{r^3}+\,\frac{\lambda\beta\gamma}{r^4}+\,\frac{\lambda^2\gamma^2}{r^5}\right).
\end{align}  \label{eq:Ch3VpscalingsS}
\end{subequations}
The algebraically growing terms in (\ref{eq:Ch3VpscalingsS}a) will be dominated by scales of $O(H)$, and the resulting contributions to the lift velocity are obtained by cutting off the divergence\,(for $r\rightarrow\infty$) at $r\sim O(\lambda^{-1})$; this cut-off recognizes the wall-induced screening of the unbounded-domain behavior that eventually leads to a more rapid decay for $r \gg O(\lambda^{-1})$, and thence, convergence. The algebraically decaying terms in both (\ref{eq:Ch3VpscalingsS}a) and (\ref{eq:Ch3VpscalingsS}b) will be dominated by scales of $O(a)$, with the lift velocity contributions now obtained by cutting off the divergence\,(for $r\rightarrow 0$) at $r\sim O(1)$. The $\ln r$ terms in (\ref{eq:Ch3VpscalingsS}a) imply the dominance of the intermediate\,(matching) interval $1 \ll r \ll \lambda^{-1}$, leading, in principle, to contributions of $O(\beta^2\ln\lambda^{-1})$ and $O(\lambda^2\gamma^2\ln \lambda^{-1})$ to the lift velocity; logarithmically smaller contributions of $O(\beta^2)$ and $O(\lambda^2\gamma^2)$ must arise from both scales of $O(a)\,(r\sim O(1))$ and $O(H)\,(r \sim O(\lambda^{-1})$). However, contributions from the inner and matching regions turn out to be zero by symmetry. Owing to the absence of walls at leading order, the $O(\beta^2)$ and $O(\beta^2\ln \lambda^{-1})$ inner and matching-region contributions must correspond to the lift on a neutrally buoyant sphere\,(or the equivalent point-particle singularity) in an unbounded simple shear flow;  likewise, the $O(\lambda^2\gamma^2)$ and $O(\lambda^2\gamma^2\ln \lambda^{-1})$ inner and matching-region contributions must correspond to the lift on a sphere at the origin of an unbounded quadratic flow. In both these scenarios, however, the two lateral directions are equivalent, and there can be no lift. That these contributions are zero may also be seen from the fact that one cannot construct a true vector, the inertial lift velocity, from any quadratic combination of the velocity gradient tensor associated with an ambient linear flow\,(simple shear for the present case of an ambient Poiseuille flow), or from any quadratic combination of the third order tensor that would characterize a generic quadratic flow. Crucially, the $O(\beta^2)$ and $O(\lambda^2\gamma^2)$ outer-region contributions are not subject to the above symmetry-argument-based limitation. This is due to the importance of walls at leading order, and the implied availability of an additional vector\,(the unit normal characterizing the wall orientations) to construct the lift velocity vector.

Based on the above arguments, one is led to the following lift velocity contributions from the linear and nonlinear inertial terms in the integrand:
\begin{subequations}
	\begin{align}
	&V_p^\text{linear}\sim Re_p\left[\beta^2(\text{outer})+\beta\gamma(\text{outer})+\lambda\gamma\beta(\text{inner})+\lambda^2\gamma^2(\text{outer})\right],\\
	&V_p^\text{non-linear}\sim Re_p\,\lambda\beta\gamma(\text{inner}).
	\end{align} \label{eq:Ch3Vpscalings2S}
\end{subequations}
From (\ref{eq:Ch3Vpscalings2S}a) and (\ref{eq:Ch3Vpscalings2S}b), the leading order contribution to the lift velocity is seen to come from the linearized inertial terms, and from scales of $O(H)$, with there being two such contributions: one proportional to $\beta^2$ that characterizes wall-induced repulsion in an ambient linear flow, and the other proportional to $\beta\gamma$ that denotes the contribution due to the ambient profile curvature. The dominance of  scales of $O(H)$ implies that, for purposes of evaluating the above contributions, the sphere in both the actual and test problems can be replaced by the corresponding point singularity. Thus, at leading order, one only need consider the terms in (\ref{eq:Ch4RTliftvelS}) that are linear in $\bm{u}_s$, and further, $\bm{u}_s$ and $\bm{u}$ may be approximated as the disturbance fields due to a stresslet ($\bm{u}_\text{str}$) and Stokeslet ($\bm{u}_\text{St}$), respectively. Note that $\bm{u}_\text{St}$ and $\bm{u}_\text{str}$ are a combination of the unbounded domain components and 
additional wall-image contributions, both of which are of comparable importance on scales of $O(H)$; detailed expressions are given in \cite{anandJeffAvgd2022S}. Thus, the inertial lift velocity at leading order in $Re_p$ and $\lambda$ reduces to:
\begin{align}
V_p &=-Re_p\int_{V^F+V^P}\bm{u}_\text{St}\cdot\left(\bm{u}_\text{str}\cdot\bm{\nabla U}^\infty+\bm{U}^\infty\cdot\bm{\nabla u}_\text{str}\right)\,dV, \label{eq:Ch3VpOuterS}
\end{align}
where the $O(\lambda\gamma)$ term in $\bm{u}^\infty$ has been neglected, so $\bm{U}^\infty =(\beta r_2+\lambda\gamma r_2^2)\bm{1}_1$ in (\ref{eq:Ch3VpOuterS}). Further, on account of the subdominance of scales of $O(a)$, the domain of integration has been extended to include the particle volume\,($V^P$), with an accompanying change in the integration variable that is now the position vector scaled by $H$\,(rather than $a$ as in (\ref{eq:Ch4RTliftvelS})). \eqref{eq:Ch3VpOuterS} is the point-particle approximation for the lift velocity for $Re_c\ll1$, and corresponds to a dimensional lift velocity of $O(V_\text{max}^2\lambda^2a/\nu)$. The use of a rescaled\,(with $\lambda$) integration variable, as mentioned above, leads to the integral in (\ref{eq:Ch3VpOuterS}) being only a function of $s$; the detailed evaluation of this integral, via a partial Fourier transform, is discussed in \cite{anandJeffAvgd2022S}.

For $Re_c \gtrsim O(1)$, the scaling arguments used above to establish the dominance of the outer region contributions still hold. The outer region now corresponds to scales of order the inertial screening length or larger, so that the farfield Stokesian estimate for $\bm{u}_s$, used above to establish outer-region dominance, remains valid only for $1 \ll r \ll \lambda^{-1}Re_c^{-\frac{1}{2}}$, with the disturbance velocity field in the actual problem decaying more rapidly for larger $r$. The outer-region dominance for $Re_c \gtrsim O(1)$ implies that this disturbance field may still be approximated as being driven by a stresslet forcing, although the forcing appears in the linearized Navier-Stokes equations. While there still exist physically distinct contributions arising from profile curvature and wall-induced repulsion, the lift velocity can no longer be written as an additive superposition of the two. Furthermore, despite the relevance of the inertial screening length, asymptotically larger scales of $O(H)$ continue to be relevant. The ratio of these two outer scales involves $Re_c$ which appears in the governing linearized equations of motion. Thus, the version of the reciprocal theorem integral in (\ref{eq:Ch3VpOuterS}) for $Re_c \gtrsim O(1)$, with $\bm{u}_s$ replaced by its finite-$Re_c$ analog, will be a function of both $s$ and $Re_c$. Although the actual calculation is more easily accomplished via a direct solution of the partially Fourier transformed ODE's using a shooting method\,\citep{schonberghinch1989S,anandJeffAvgd2022S}, one may nevertheless write the point-particle lift velocity contribution formally in the form $(V_\text{max}^2\lambda^2 a/\nu)F_1(s;Re_c)$.

The modification of the leading order lift velocity, for $\lambda$ small but finite, arises from contributions in (\ref{eq:Ch3Vpscalings2S}) that are of a smaller order in $\lambda$ than the outer-region contributions included in (\ref{eq:Ch3VpOuterS}). The largest such contributions are proportional to $\lambda(\beta\gamma)$, and pertain to the inner region, arising from both the linear and nonlinear inertial terms in (\ref{eq:Ch3Vpscalings2S}a) and (\ref{eq:Ch3Vpscalings2S}b). The $\beta\gamma$-dependence implies that these contributions arise solely due to the coupling of the shear and curvature of the ambient profile, consistent with earlier symmetry arguments. Further, scales of $O(H)$ are irrelevant, implying that the $\lambda(\beta\gamma)$ terms will lead to contributions independent of $Re_c$, with the dimensional lift velocity being of the form $(V_\text{max}^2\lambda^3 a/\nu)F_2(s)$, as in equation (2) of the main manuscript. A second implication of the $O(\lambda\beta\gamma)$ contributions being from the inner region is that they arise independently of the outer-region point-particle contribution. Said differently, the $(V_\text{max}^2\lambda^2a/\nu)F_1(s;Re_c)$ and $(V_\text{max}^2\lambda^3a/\nu)F_2(s)$ contributions to the inertial lift correspond to the leading order terms of the underlying asymptotic expansions of the integrand in the outer and inner regions, respectively. This implies that the  $O(V_\text{max}^2\lambda^3 a/\nu)F_2(s)$ contribution is not a correction to the leading order point-particle result, and thereby, not constrained to be small in comparison. This feature is especially significant since the emergence of multiple equilibria in the lift profiles\,(Region \textcircled{2} in Fig\,4 of the main manuscript) is only made possible by allowing the finite-size contribution to be comparable to the leading point-particle one.
 
Note that there are other corrections for finite $\lambda$: for instance, the $O(\lambda^2\gamma^2)$ outer-region contribution in (\ref{eq:Ch3Vpscalings2S}), the correction to $\bm{u}_s$ arising from $\lambda$-dependent corrections to the stresslet coefficient, etc. While both of these may be shown to be of a smaller order in $\lambda$ for $Re_c \ll 1$, they will be far smaller for large $Re_c$ owing to an overall weakening of wall-induced corrections. This weakening arises from the more rapid decay of the disturbance velocity field for distances larger than $O(HRe_c^{-\frac{1}{2}})$. This $Re_c$-dependence is important since, as explained in the main manuscript, the finite-size contributions become significant only at large $Re_c$ owing to the reduction in the magnitude of the point-particle contribution with increasing $Re_c$. Clearly, the only finite-size contributions of relevance correspond to the $O(\lambda \beta\gamma)$ terms in (\ref{eq:Ch3Vpscalings2S}a) and (\ref{eq:Ch3Vpscalings2S}b). To calculate these, we return to the original reciprocal theorem result, viz.\,\eqref{eq:Ch4RTliftvelS},  and subtract and add back the leading point-particle contribution given by \eqref{eq:Ch3VpOuterS}:
\begin{align}
	V_p=&-Re_p\int_{V^F+V^P}\bm{u}_\text{St}\cdot\big(\bm{u}_\text{str}\cdot\bm{\nabla U}^\infty+\bm{U}^\infty\cdot\bm{\nabla u}_\text{str}\big)\,dV \nonumber \\
 &-Re_p\biggl[\int_{V^F}\bm{u}\cdot\big(\bm{u}_s\cdot\bm{\nabla u}_s+\bm{u}_s\cdot\nabla \bm{u}^\infty+\bm{u}^\infty\cdot\bm{\nabla u}_s\big)\,dV\nonumber\\
	&-\int_{V^F+V^P}\bm{u}_\text{St}\cdot\big(\bm{u}_\text{str}\cdot\bm{\nabla U}^\infty+\bm{U}^\infty\cdot\bm{\nabla u}_\text{str}\big)\,dV\biggr].
	\label{eq:Ch4Derivation1}
\end{align}
Splitting the point-particle contribution within brackets into separate integrals over the fluid\,($V^F$) and particle\,($V^P$) domains, separating the nonlinear and linear inertial terms in the original reciprocal theorem integral, and then combining the integrals involving the linear inertial terms, one obtains:
\begin{align}
	V_p=&-Re_p\int_{V^F+V^P}\bm{u}_\text{St}\cdot\big(\bm{u}_\text{str}\cdot\bm{\nabla U}^\infty+\bm{U}^\infty\cdot\bm{\nabla u}_\text{str}\big)\,dV \nonumber\\
 &-Re_p\biggl[ \int_{V^F}\bm{u}\cdot\big(\bm{u}_s\cdot\bm{\nabla u}_s\big)\,dV+\int_{V^F}\big[\bm{u}\cdot\big(\bm{u}_s\cdot\nabla \bm{u}^\infty+\bm{u}^\infty\cdot\bm{\nabla u}_s\big)\nonumber\\
	&-\bm{u}_\text{St}\cdot\big(\bm{u}_\text{str}\cdot\bm{\nabla U}^\infty+\bm{U}^\infty\cdot\bm{\nabla u}_\text{str}\big)\big]\,dV- \int_{V^P}\bm{u}_\text{St}\cdot\big(\bm{u}_\text{str}\cdot\bm{\nabla U}^\infty\nonumber\\
 &+\bm{U}^\infty\cdot\bm{\nabla u}_\text{str}\big)\,dV \biggr].
	& \label{eq:Ch4Derivation2S}
\end{align}
The last integral within brackets, over $V^P$, evidently involves scales of $O(a)$, and we therefore focus on the remaining two integrals. The first integral within brackets contains the nonlinear inertial term, and as already seen in (\ref{eq:Ch3VpscalingsS}b), the integrand exhibits an algebraic decay for $r \gg 1$. The second integral involves the difference between the linearized inertial terms, and their approximations based on point-particle representations of the corresponding velocity fields, as a result of which growing terms cancel out, and only the algebraically decaying one in (\ref{eq:Ch3Vpscalings2S}a) survives. The algebraic decay implies the dominance of scales of $O(a)$, and therefore, that wall-induced image contributions do not contribute at leading order. Thus, the leading approximations of all of the integrals within brackets, in (\ref{eq:Ch4Derivation2S}), may be obtained by replacing the original fluid domain\,($V^F$) by a completely unbounded one\,($V^\infty$), and (\ref{eq:Ch4Derivation2S}) may be written in the form:
\begin{align}
	V_p=&-Re_p\int_{V^F+V^P}\bm{u}_\text{St}\cdot\big(\bm{u}_\text{str}\cdot\bm{\nabla U}^\infty+\bm{U}^\infty\cdot\bm{\nabla u}_\text{str}\big)\,dV. \nonumber\\
 &+ \lambda Re_p\biggl[ -\int_{V^\infty}\bm{u}_i\cdot\big(\bm{u}_{s,i}\cdot\bm{\nabla}\bm{u}_{s,i}\big)dV-\int_{V^\infty}\big[\bm{u}_i\cdot\big(\bm{u}_{s,i}\cdot\nabla \bm{u}^\infty+\bm{u}^\infty\cdot\bm{\nabla u}_{s,i}\big)\nonumber\\
        &-\bm{u}_{\text{St},i}\cdot\big(\bm{u}_{\text{str},i}\cdot\bm{\nabla U}^\infty+\bm{U}^\infty\cdot\bm{\nabla u}^{(1)}_{\text{str},i}\big)\big]\,dV+\int_{V^P}\bm{u}_{\text{St},i}\cdot\big(\bm{u}_{\text{str},i}\cdot\bm{\nabla U}^\infty\nonumber\\
 &+\bm{U}^\infty\cdot\bm{\nabla u}_{\text{str},i}\big)\,dV \biggr],	\label{eq:Ch4InnerIntegral3S}
\end{align}
where both the actual velocity fields and their point-particle approximations, that appear in the three bracketed integrals, are now approximated by simpler expressions pertaining to an unbounded fluid domain; these are indicated by the additional subscript `\textit{i}', and are given below:: 
\begin{subequations}
	\begin{align}
		\bm{u}_i&=\frac{\bm{1}_2}{8\pi}\cdot\bigg(\frac{\bm{I}}{r}+\frac{\bm{rr}}{r^3}\bigg)+\frac{\bm{1}_2}{8\pi}\cdot\bigg(\frac{\bm{I}}{3r^3}-\frac{\bm{rr}}{r^5}\bigg), \label{utest:innerS}\\
		\bm{u}_{s,i}&=-\frac{5}{2}\frac{(\bm{E}:\bm{rr})\bm{r}}{r^5}- \bigg[\frac{\bm{E}\cdot\bm{r}}{r^5}-\frac{5 (\bm{E}:\bm{rr})\bm{r}}{2r^7}\bigg]\nonumber\\
  &+\lambda\gamma\big\{\!\!-\!\frac{1}{24r^9}\big[3r^6+r^4(-23r_1^2+r_2^2-8r_3^2-3)+15r^2\big(r_1^2(7 r_2^2+1)+r_2^2\big)\nonumber\\
        &-105r_1^2 r_2^2\big]\bm{1}_1+\frac{5}{8r^9}(r^2-1)r_1 r_2(3r^2-7r_2^2)\bm{1}_2+\frac{5}{8r^9}(r^2-1)r_1 r_3 (r^2-7r_2^2) \bm{1}_3 \big\}, \label{uactual:innerS} \\
		\bm{u}_{\text{St},i}&=\frac{\bm{1}_2}{8\pi}\cdot\bigg(\frac{\bm{I}}{r}+\frac{\bm{rr}}{r^3}\bigg), \label{StokesletS}\\
		\bm{u}_{\text{str},i}&=-\frac{5}{2}\frac{(\bm{E}:\bm{rr})\bm{r}}{r^5}. \label{StressletS}
	\end{align} \label{eq:Ch4uunboundedS}
\end{subequations}
(\ref{utest:innerS}) is the disturbance velocity field induced by a translating sphere. (\ref{uactual:innerS}) is the disturbance field due to a force-free sphere in an ambient flow with both linear and quadratic components: the part involving $\bm{E}\,(=\frac{\beta}{2}(\bm{1}_1\bm{1}_2+\bm{1}_2\bm{1}_1)$, the rate of strain tensor of the ambient Poiseuille flow), constitutes the disturbance in an ambient linear flow, while that proportional to $\gamma$ constitutes the response to the quadratic component. As originally shown\,(but not used) by \cite{holeal1974S}, the latter may be obtained using an expansion in spherical harmonics \citep{kimkarrilaS}. (\ref{StokesletS}) and (\ref{StressletS}) are the usual velocity fields for a Stokeslet and the stresslet in an unbounded domain.

Finally, note that the integrals over $V^P$ and $V^\infty$ extend down to the origin\,(the sphere center) with both integrands including an $O(\beta^2)$ contribution arising from the coupling of the linear ambient flow component with the stresslet field, that is $O(1/r^3)$ for $r \rightarrow 0$\,(see initial integrand estimates); while this leads to a conditionally convergent behavior, doing the angular integration first leads to a trivial answer, as must be the case based on symmetry arguments above. For the integral over $V^\infty$, a similar conditionally convergent behavior arises at infinity from the leading $O(\gamma^2)$ contribution, that is again resolved by appropriate choice of the order of integration. Even the leading point-particle integral is conditionally convergent at the origin, an issue (implicitly)\,addressed by a particular order of integration in the calculation procedure\citep{anandJeffAvgd2022S}.
%\begin{align}	V_p=&-Re_p\int_{V^F+V^P}\big[\bm{u}_\text{St}\cdot\big(\bm{u}_\text{str}\cdot\bm{\nabla U}^\infty+\bm{U}^\infty\cdot\bm{\nabla u}_\text{str}\big)-\bm{u}_{\text{St},i}\cdot\big(\bm{u}_{\text{str},i}\cdot\bm{\nabla}(\beta r_2 \bm{1}_1)\nonumber\\	&+\beta r_2 \bm{1}_1\cdot\bm{\nabla u}_{\text{str},i}\big)\big]\,dV+\lambda Re_p\Bigg[-\int_{V^\infty}\bm{u}_i\cdot\big(\bm{u}_{\beta,i}\cdot\bm{\nabla}\bm{u}_{\gamma,i}+\bm{u}_{\gamma,i}\cdot\bm{\nabla}\bm{u}_{\beta,i}\big)\,dV\nonumber\\    &-\int_{V^\infty}\Big\{\big[\bm{u}_i\cdot\big(\bm{u}_{s,i}\cdot\nabla \bm{u}^\infty+\bm{u}^\infty\cdot\bm{\nabla u}_{s,i}\big)-\bm{u}_{\text{St},i}\cdot\big(\bm{u}_{\text{str},i}\cdot\bm{\nabla U}^\infty\nonumber\\    &+\bm{U}^\infty\cdot\bm{\nabla u}_{\text{str},i}\big)\big]-\Big[\bm{u}_{\text{St},i}\cdot\big(\bm{u}_{q,i}\cdot\bm{\nabla}(\gamma r_2^2\bm{1}_1)+\gamma r_2^2\bm{1}_1\cdot\bm{\nabla u}_{q,i} \big)\Big]\Big\}\,dV\nonumber\\ &+\int_{V^P}\big[\bm{u}_{\text{St},i}\cdot\big(\bm{u}_{\text{str},i}\cdot\bm{\nabla U}^\infty+\bm{U}^\infty\cdot\bm{\nabla u}_{\text{str},i}\big)-\bm{u}_{\text{St},i}\cdot\big(\bm{u}_{\text{str},i}\cdot\bm{\nabla}(\beta r_2 \bm{1}_1)\nonumber\\    &+\beta r_2 \bm{1}_1\cdot\bm{\nabla u}_{\text{str},i}\big)\big]\,dV\Bigg]	\label{eq:Ch4VpIntegralFinal}\end{align}
%Here, $\bm{u}_{q,i}$ is obtained from taking the large-$r$ limit of $\bm{u}_{\gamma,i}$ in (\ref{uactual:gammainner}).

The volume integrals, within brackets in \eqref{eq:Ch4uunboundedS}, are readily calculated analytically:
\begin{align}	
\int_{V^\infty}\bm{u}_i\cdot\big(\bm{u}_{s,i}\cdot\bm{\nabla}\bm{u}_{s,i}\big)\,dV\!=&\frac{6143\,\beta\gamma}{120960}, \label{result:1S}\\	\hspace*{-1.in}\int_{V^\infty}\!\!\big[\bm{u}_i\!\cdot\!\big(\bm{u}_{s,i}\!\cdot\!\nabla \bm{u}^\infty\!+\!\bm{u}^\infty\!\cdot\!\bm{\nabla u}_{s,i}\big)\!-\!\bm{u}_{\text{St},i}\!\cdot\!\big(\bm{u}_{\text{str},i}\!\cdot\!\bm{\nabla U}^\infty\!+\!\bm{U}^\infty\!\cdot\!\bm{\nabla u}_{\text{str},i}\big)\big]dV\!\!=&\frac{37\,\beta\gamma}{1260},\label{result:2S}\\	\int_{V^P}\bm{u}_{\text{St},i}\cdot\big(\bm{u}_{\text{str},i}\cdot\bm{\nabla U}^\infty+\bm{U}^\infty\cdot\bm{\nabla u}_{\text{str},i}\big)\,dV=&\!-\frac{\beta\gamma}{4}. \label{result:3S}
 \end{align}
Substituting (\ref{result:1S}-\ref{result:3S}) in \eqref{eq:Ch4Derivation2S}, with $\beta=4(1-2s)$ and $\gamma=-4$, gives:
\begin{align}
	V_p=Re_p\big[F_1(s,Re_c)+\lambda F_2(s)\big],
	\label{eq:Ch4VpFinalS}
\end{align}
with $F_2(s)=\frac{1141(1-2s)}{216}$.
%%%%%%%%%%%%%%%%%%%%%%%%%%%%%%%%%%%%%%%
%\bibliographystyle{jfm}
%\bibliography{references}

\begin{thebibliography}{60}
\bibitem{footnote1} $Re_c$ will be used here to denote the Reynolds number based on the relevant macroscopic scale, either the channel width or the pipe diameter.
\bibitem{footnote5} The SS-equilibria for drops, within the framework of a point-particle formulation, may be shown to be identical to those for rigid spheres. The point-particle disturbance velocity field depends only on the strength of the induced stresslet, and as a result, the inertial lift for a spherical drop may be obtained from that for a rigid sphere by multiplication by the ratio of the corresponding stresslet coefficients.
\bibitem{footnote6}  Wherever appropriate, $Re_c$ is taken to denote the Reynolds number based on the pipe diameter and the mean velocity of the flow through the pipe.
\end{thebibliography}

\end{document}